\documentclass[aps,preprintnumbers,prl,twocolumn,superscriptaddress]{revtex4} 

\usepackage{slashed}
\usepackage{graphicx}
\usepackage{color}
\usepackage{amsfonts}

\newcommand{\gsim}{\buildrel > \over {_\sim}}

\newcommand{\be}{\begin{equation}}
\newcommand{\ee}{\end{equation}}
\newcommand{\bee}{\begin{equation*}}
\newcommand{\eee}{\end{equation*}}
\newcommand{\bea}{\begin{eqnarray}}
\newcommand{\eea}{\end{eqnarray}}
\newcommand{\bean}{\begin{eqnarray*}}
\newcommand{\eean}{\end{eqnarray*}}
\newcommand{\met}{{E\!\!\!\slash}_T}

\begin{document}

\preprint{ACFI-T15-11}

\title{TeV Lepton Number Violation: From Neutrinoless Double $\beta$-Decay to the LHC}

\author{Tao Peng}
\email{tpeng23@wisc.edu}
\affiliation{Department of Physics, University of Wisconsin-Madison, Madison, WI 53706}
\author{Michael J. Ramsey-Musolf}
\email{mjrm@physics.umass.edu}
\affiliation{Amherst Center for Fundamental Interactions, Department of Physics, University of Massachusetts-Amherst, Amherst, MA 01003 USA}
\affiliation{Kellogg Radiation Laboratory, California Institute of Technology, Pasadena, CA 91125, USA}
\author{Peter Winslow}
\email{pwinslow@physics.umass.edu}
\affiliation{Amherst Center for Fundamental Interactions, Department of Physics, University of Massachusetts-Amherst, Amherst, MA 01003 USA}

\begin{abstract}
We analyze the sensitivity of next-generation tonne-scale neutrinoless double $\beta$-decay ($0\nu\beta\beta$) experiments and  searches for same-sign di-electrons plus jets at the Large Hadron Collider to TeV scale lepton number violating interactions. Taking into account previously unaccounted for physics and detector backgrounds at the LHC, renormalization group evolution, and long-range contributions to $0\nu\beta\beta$ nuclear matrix elements, we find that the reach of tonne-scale $0\nu\beta\beta$ generally exceeds that of the LHC. However, for a range of heavy particle masses near the TeV scale, the high luminosity LHC and tonne-scale $0\nu\beta\beta$ may provide complementary probes.
\end{abstract}

\maketitle
Total lepton number (L) is a conserved quantum number at the classical level in the Standard Model (SM) of particle physics, yet it is not conserved in many scenarios for physics beyond the Standard Model (BSM). Among the most widely considered classes of BSM scenarios admitting lepton number violation (LNV) is the see-saw mechanism\cite{Minkowski:1977sc,Mohapatra:1979ia,Glashow:qal,GellMann:1980vs,Yanagida:1979exq} for neutrino mass generation that postulates the existence of heavy right-handed Majorana neutrinos, $N_R$. The associated mass operator violates L by two units, and the coupling of the $N_R$ with the left-handed SM lepton-doublet induces a corresponding L-violating Majorana mass for the three lightest neutrino mass eigenstates. Experimentally, the observation of neutrinoless double beta-decay ($0\nu\beta\beta$-decay) of atomic nuclei would provide direct evidence for LNV. This observation would also indicate the existence of a Majorana mass term for the lightest neutrinos\cite{Schechter:1981bd}, consistent with the prediction of the see-saw mechanism.

Recent results from the EXO\cite{Albert:2014awa}, GERDA\cite{Agostini:2013mzu}, and KamLand-ZEN\cite{Gando:2012zm,Asakura:2014lma} experiments have placed stringent upper limits on the $0\nu\beta\beta$-decay half lives ($T_{1/2}^{0\nu\beta\beta}$) of $^{76}$Ge and $^{136}$Xe on the order of a  few times $10^{25}$ years. When interpreted in terms of the exchange of light Majorana neutrinos, these limits imply an upper bound of order 100-400 meV on the $0\nu\beta\beta$-decay effective mass $m_{\beta\beta}$, depending on the value of the nuclear matrix element employed in this extraction\footnote{A signal for the $0\nu\beta\beta$-decay of $^{76}$Ge has been reported in Ref.~\cite{KlapdorKleingrothaus:2006ff}, though it remains controversial and in conflict with the null result obtained by the GERDA collaboration.}. The next generation of \lq\lq tonne scale" $0\nu\beta\beta$-decay searches aim for half life sensitivities of order $\gsim 10^{27}$ years, with a corresponding $m_{\beta\beta}$ sensitivity on the order of tens of meV, consistent with expectations based on the inverted hierarchy (IH) for the light neutrino mass spectrum. In this interpretive framework, a null result would imply that either neutrinos are Majorana particles with a mass spectrum in the normal hierarchy (NH) or that they are Dirac fermions.

It is possible that neutrino oscillation studies may determine the neutrino mass hierarchy before the next generation $0\nu\beta\beta$-decay searches reach their goal sensitivity. Should the hierarchy turn out to be normal, for which $m_{\beta\beta}$  may range anywhere from the present upper bound to a magnitude $ << 10$ meV, a null result from the tonne-scale $0\nu\beta\beta$-decay experiments would not be surprising. However, alternate decay mechanisms could still lead to observation of a signal in the next generation searches, even if the light neutrino spectrum follows the NH and the value of $m_{\beta\beta}$ is experimentally inaccessible. These mechanisms include radiative neutrino mass scenarios\cite{Ma:1998dn}
and the low-scale see-saw mechanism\cite{Wyler:1982dd,Mohapatra:1986bd,Nandi:1985uh,Branco:1988ex,Pilaftsis:1991ug}\footnote{The presence of an additional \lq\lq sterile" neutrino that mixes with the three \lq\lq active" neutrinos could yield a value of $m_{\beta\beta}$ within reach of the tone-scale searches}. In these scenarios, the LNV interactions may involve particles whose masses are of order one TeV and whose exchange generates short range interactions that lead to the $0\nu\beta\beta$-decay.
Straightforward arguments indicate that the resulting $0\nu\beta\beta$-decay half-life  can be of order $10^{27}$ yr or shorter, comparable to
expectations based on the three light Majorana exchange mechanism and the IH\cite{Cirigliano:2004tc}. The associated light Majorana masses may nevertheless follow the NH with $m_{\beta\beta}$ well below the meV scale.

\begin{table*}[t!]
\centering
\begin{tabular}{||c|c|c|c|c|c|c|c|c|c|c|c||}
\hline
${\mathbb \sigma}$\bf{(fb)} & \bf{Signal} & \multicolumn{9}{|c|}{\bf{Backgrounds}} & $\mathbf{\frac{S}{\sqrt{S+B}}} \; (\sqrt{\text{{\bf fb}}})$\\
 \hline
 & & \multicolumn{3}{|c|}{Diboson} & \multicolumn{2}{|c|}{Charge Flip} & \multicolumn{4}{|c|}{Jet Fake} & \\
 \hline
 &  & W$^-$W$^-$+2j & W$^-$Z+2j & ZZ+2j & $Z/\gamma^*$+2j & $t \overline{t}$ & $t \overline{t}$ & $\overline{t}$+3j & $W^-$+3j & 4j  & \\
\hline
Before Cuts & 0.142 & 0.541 & 6.682  & 0.628 & 903.16 & 68.2 & 6.7 & 0.45 & 15.09 &  362.352 & 0.0038 \\
\hline
Signal Selection & 0.091 & 0.358 & 4.66 & 0.435 & 721.7 & 28.9 & 2.37 & 0.22 & 11.73 & 72.03 & 0.0031 \\
\hline
$H_T (\text{jets})> 650$ GeV & 0.054 & 0.04 & 0.187 & 0.015 & 5.6 & 0.266 & 0.025 & 0.0003 & 0.102 & 0.027 & 0.0213 \\
\hline
$m_{\ell_1 \ell_2}> 130$ GeV & 0.039 & 0.029 & 0.105 & 0.008 & 0.163 & 0.127 & 0.024 & 3x10$^{-4}$ & 0.101 & 0.027 & 0.0493 \\
\hline
$\met < 40$ GeV & 0.036 & 0.005 & 0.036 & 0.007 & 0.126 & 0.014 & 0.005 & 3x10$^{-5}$ & 0.03 & 0.017 & 0.0684 \\
\hline
($\eta_{j_{1,2}} - \eta_{\ell_{1,2}})_{max} < 2.2$ & 0.033 & 0.003 & 0.022 & 0.005 & 0.093 & 0.009 & 0.004 & 2x10$^{-5}$ & 0.019 & 0.011 & 0.0738 \\
\hline
\end{tabular}
\caption{{ Cut-flow designed for optimizing signal relative to background. Note: kinematic cuts are not commutative. }}
\label{CutFlow}
\end{table*}
How might one experimentally distinguish the TeV LNV scenario for $0\nu\beta\beta$-decay from the more conventional paradigm based solely on the exchange of light Majorana neutrinos? One possibility is to analyze experiments that search for charged lepton flavor violation, as discussed in Ref.~\cite{Cirigliano:2004tc}. Another, perhaps more direct, means is to search for the LNV interactions in high energy collider experiments (see, {\em e.g.} \cite{Keung:1983uu,Perez:2008ha,Melfo:2011nx,Berger:2013sir,Aad:2011vj,Chatrchyan:2012paa,Chatrchyan:2012ira,ATLAS:2012hpa,ATLAS:2012sna,ATLAS:2012ai,TheATLAScollaboration:2013jha}). 

This possibility has recently been explored by the authors of Refs.~\cite{Helo:2013dla,Helo:2013ika}, who utilized a simplified model framework to analyze the relative sensitivities of tonne-scale $0\nu\beta\beta$-decay experiments and searches for LNV signals at the CERN Large Hadron Collider. These authors performed a systematic classification of simplified models that one may map onto more complete theories, such as   R-parity violating supersymmetry. They find that in a broad range of cases the LHC with 300 fb$^{-1}$ of integrated luminosity (corresponding to the end of Run II) would achieve substantially greater reach for TeV-scale LNV interactions than would the tonne-scale $0\nu\beta\beta$-decay searches\footnote{This conclusion assumes less than three signal events with 300 fb$^{-1}$.}. If verified, the prospective LHC exclusion of TeV scale LNV, coupled with identification of the NH for the light neutrino mass spectrum, could render the prospects for discovery with tonne-scale $0\nu\beta\beta$-decay searches less compelling than presently considered.

In what follows, we revisit the analysis of Refs.~\cite{Helo:2013dla,Helo:2013ika} and find that their conclusions regarding the LHC reach may be overly optimistic. We consider three aspects of the LHC and $0\nu\beta\beta$-decay physics not included in Refs.~\cite{Helo:2013dla,Helo:2013ika}: (a) the impact of SM and detector backgrounds on the significance of an LHC LNV signal; (b) running of the corresponding LNV effective operators from the TeV scale to the low-energy scale relevant to  $0\nu\beta\beta$-decay; and (c) long-distance contributions to the $0\nu\beta\beta$-decay nuclear matrix element (NME). The impacts of these considerations are, respectively, to (a) degrade the significance of the LHC LNV signal for a given choice of LNV model parameters; (b) reduce the strength of the $0\nu\beta\beta$-decay amplitude relative to the inferred value of parameters at the high scale; and (c) enhance the NME. We then find that for a limited range of heavy particle masses, existing $0\nu\beta\beta$-decay searches and Run II of the LHC may have comparable sensitivities to TeV scale LNV, depending on the values of the $0\nu\beta\beta$-decay nuclear and hadronic matrix elements. Accumulation of additional data with the high-luminosity phase of the LHC would be necessary to achieve a reach comparable to the tonne-scale $0\nu\beta\beta$-decay searches.

To be concrete, we focus on  one of the simplified models yielding the greatest LHC reach according to Refs.~\cite{Helo:2013dla,Helo:2013ika}. The model includes a scalar doublet $S$ transforming as $(1, 2, 1)$ under SU(3$)_C\times$SU(2$)_L\times$U(1$)_Y$ and a Majorana fermion $F$ that transforms as a SM gauge singlet. The interaction Lagrangian is
\be
\mathcal{L}_\mathrm{LNV} = g_1 {\bar Q}^{\alpha}_i d^\alpha_i S + g_2 \epsilon^{ij} {\bar L}_i F S^\ast_j +\mathrm{h.c.}\ \ \ ,
\label{eq:LNV}
\ee
where $L$ and $Q$ are first generation left-handed lepton and quark doublets, respectively; $d$ is the right-handed down quark; and Roman and Greek indices correspond to SU(2$)_L$ and SU(3$)_C$ components, respectively. In high energy proton-proton collisions, the interaction (\ref{eq:LNV}) will generate a final state with a same sign (SS) di-electron pair along with two high-$p_T$ jets. When either the $S$ or $F$ appears as an $s$-channel resonance, the corresponding cross section will be enhanced. For the low-energy $0\nu\beta\beta$-decay process, one may integrate out the heavy degrees of freedom, yielding the dimension-nine LNV interaction:
\be
\label{eq:LNV2}
\mathcal{L}_\mathrm{LNV}^\mathrm{eff} = \frac{C_1}{\Lambda^5} \mathcal{O}_1+\mathrm{h.c.}\ \ \ , \quad \mathcal{O}_1 = {\bar Q} \tau^+d {\bar Q} \tau^+ d {\bar L} L^C\ \ \ ,
\ee
where  $L^C$ is the lepton doublet charge conjugate field, $C_1=g_1^2 g_2^2$ and $\Lambda^5 = M_S^4 M_F$. 

We have implemented the model (\ref{eq:LNV}) in Madgraph and generated events with Madevent~\cite{Alwall:2014hca} for $pp$ collisions at 14 TeV, carrying out showering, jet matching, and hadronization with Pythia~\cite{Sjostrand:2007gs}	 and detector simulation with PGS. The dominant backgrounds involve (a) \lq\lq charge flip", wherein one lepton from a SM opposite sign (OS) di-electron pair transfers most of its $p_T$ to an electron of the opposite sign through conversion and (b) a high-$p_T$ jet is registered as an electron in the electromagnetic calorimeter (\lq\lq jet fake"). The largest contributors to the charge flip background are SM production of a $Z$ and virtual $\gamma$ plus jets, followed by $t{\bar t}$ production wherein the $b$-quarks from the top decays are not tagged. For the jet fake background, SM multi-jet production is by far the leading contributor. Subdominant backgrounds include diboson (WW, WZ, ZZ) plus jets. The charge flip background from the various aforementioned sources was derived by binning events in pseudo-rapidity ($\eta$) and applying the $\eta$-dependent charge-flip probabilities as measured by ATLAS~\cite{Aad:2011mk}. For the jet-fake background, we applied a medium jet-fake probability of $2\times 10^{-4}$ \cite{Aad:2011mk,Carlo} times a combinatoric factor associated with the number of jet-fakes in an event with $N$ jets.

After imposing a set of basic selection cuts
($
p_{T_{j,b,\ell^\pm}} > 20 \text{ GeV}, \quad | \eta_j | < 2.8, \quad | \eta_{\ell^\pm} | < 2.5 
$)
we find that additional cuts on $H_T(\mathrm{jets})$,  the scalar sum of all jet $p_T$, the dilepton invariant mass, and missing energy $\met$ are highly effective in reducing the background while maintaining the signal. A  set of cuts that optimizes the significance $S/\sqrt{S+B}$ is given in Table \ref{CutFlow}. The signal indicated is generated for $M_S=M_F=1$ TeV and $g_1 = g_2 = 0.176$, corresponding to a $0\nu\beta\beta$-decay rate consistent with the present GERDA upper bound (see below). 


In order to translate the sensitivity to the parameters that enter the high energy process to the $0\nu\beta\beta$-decay rate, we evolve the operator $\mathcal{O}_1$ to the GeV scale using the renormalization group. Between the scale $\mu=\Lambda$ and the weak scale $v=246$ GeV, one must in principle include both QCD and electroweak corrections. As the latter are generally considerably smaller than the former, we include only the QCD corrections and continue the running from $v$ to $\mu=1$ GeV. Under this evolution, $\mathcal{O}_1$ will mix with three additional operators:
$\mathcal{O}_2 = {\bar Q}\sigma_{\mu\nu} \tau^+ d {\bar Q}\sigma^{\mu\nu} \tau^+d {\bar L} L^C$,
$\mathcal{O}_3 = {\bar Q}T^A \tau^+ d {\bar Q}T^A \tau^+ d {\bar L} L^C$, and 
$\mathcal{O}_4 = {\bar Q}\sigma_{\mu\nu} T^A\tau^+ d {\bar Q}\sigma^{\mu\nu} T^A \tau^+ d {\bar L} L^C $, 
where $T^A$ $A=1,\cdots 8$ denote the SU(3$)_C$ generators in the fundamental representation and $\tau^+$ is the isospin raising operator. The corresponding anomalous dimension matrix is
\be
\gamma^T =\frac{\alpha_s}{2\pi} \left(
\begin{array}{cccc}
-8 & 0 & 0 & -32/3\\
0 & -8/3 & 2/9 & 0 \\
0 & - 48 & 1 & -20\\
-1 & 0 & -5/12 & -19/3
\end{array}\right)
\ \ \ .
\ee
The  Wilson coefficients $C^T=(C_1, \cdots, C_4)$ then evolve according to $dC/d\ln\mu = \gamma^T C$.
Under this evolution, we find, for example, that if only $C_1(\mu=\Lambda)$ is non-vanishing at the high scale, then the magnitude of the Wilson coefficients  $C_j(\mu=1\ \mathrm{GeV})$ are: 
$C_1=0.203 C_1(\Lambda)$, $C_2 =-0.007C_1(\Lambda)$, $C_3=0.266C_1(\Lambda)$, and $C_4=-0.055C_1(\Lambda)$. 

For $\mu$ below $\sim 1$ GeV, use of quark degrees of freedom is no longer appropriate, so one must match the operators $\mathcal{O}_j$ onto operators built from hadronic degrees of freedom. To that end, we follow Ref.~\cite{Prezeau:2003xn} and exploit the transformation properties of the $\mathcal{O}_j$ under SU(2$)_L\times$SU(2$)_R$ chiral symmetry.  It is convenient to Fierz transform $\mathcal{O}_{3,4}$ to forms in which all quark blinears are color singlets, leading to an effective coefficient of $\mathcal{O}_1$ given by 
\be
C_\mathrm{eff} \approx C_1(1\ \mathrm{GeV}) -\frac{5}{12}C_3(1\ \mathrm{GeV})  = 0.092 C_1(\Lambda)
\nonumber
\ee
where we have omitted the negligible contributions from $C_{2,4}(1\ \mathrm{GeV})$. Using the notation of Ref.~\cite{Prezeau:2003xn} we note that the part of $\mathcal{O}_1$ relevant to the decay process is
\be
\mathcal{L}_\mathrm{LNV}^\mathrm{eff} = \frac{C_\mathrm{eff}}{2\Lambda^5}\left(\mathcal{O}_{2+}^{++} -\mathcal{O}_{2-}^{++}\right){\bar e_L} e_R^c\ + \mathrm{h.c.} \ \ \ ,
\label{eq:loweop}
\ee
where $e_R^c\equiv (e_L)^C$ and 
\be
\mathcal{O}_{2\pm}^{ab} = {\bar q}_R\tau^a q_L {\bar q}_R \tau^b q_L \pm {\bar q}_L\tau^a q_R {\bar q}_L \tau^b q_R
\ee
with $q_{L,R}^T=(u,d)_{L,R}$. Since $\mathcal{O}_{2-}^{++}$ is parity-odd and the $0\nu\beta\beta$-decay processes of experimental interest involve $0^+\to 0^+$ transitions, we retain only the $\mathcal{O}_{2+}^{++}$ part of  (\ref{eq:loweop}).

At the hadronic level, $\mathcal{O}_{2+}^{++} {\bar e_L} e_R^c$ matches onto the two pion-two electron operator
\be
\label{eq:pioneff}
\frac{C_\mathrm{eff}}{\Lambda} \mathcal{O}_{2+}^{++}{\bar e_L} e_R^c +\mathrm{h.c.} \to \frac{C_\mathrm{eff}\Lambda_H^2 F_\pi^2}{2\Lambda^5} \pi^-\pi^- {\bar e_L} e_R^c +\mathrm{h.c.}\, ,
\ee
where $F_\pi=92.2\pm 0.2$ MeV  is the pion decay constant \cite{DGH:14} and $\Lambda_H$ is a mass scale associated with hadronic matrix elements of the four quark operator $\mathcal{O}_{2+}^{++}$.  Using the vacuum saturation and factorization approximation, we estimate the latter to be
$ \Lambda_H = m_\pi^2/(m_u+m_d) \approx 2.74$ GeV for $m_{\pi^+}= 139$ MeV and $m_u+m_d=7$ MeV  \cite{Agashe:2014kda}.

The effective pion-electron interaction in Eq.~(\ref{eq:pioneff}) leads to a long-range contribution to the $0\nu\beta\beta$ amplitude\cite{Prezeau:2003xn}. Following Ref.~\cite{Prezeau:2003xn} we then obtain the following result for the decay rate:
\bea
\label{eq:rate}
\frac{1}{T_{1/2}} & = & \left[G_{0\nu}\times (1\, \mathrm{TeV})^2\right] \left(\frac{\Lambda_H}{\mathrm{TeV}}\right)^4 \left(\frac{1}{18}\right) \\
&\times&
\nonumber
\left(\frac{v}{\mathrm{TeV}}\right)^8
\left(\frac{1}{\cos\theta_C}\right)^4 \vert M_0\vert^2 \, \left[\frac{C_\mathrm{eff}^2}{(\Lambda/\mathrm{TeV})^{10}}\right]\ \ \ , \\
\nonumber
G_{0\nu} &=& (G_F \cos\theta_C)^4 \left(\frac{\hbar c}{R}\right)^2 \left(\frac{1}{32\pi^2 \hbar\ln 2}\right) I(E_{\beta\beta})\ \ \ ,
\eea
with $\theta_C$ being the Cabibbo angle, $I(E_{\beta\beta})$ the electron phase space integral
\be
 \int_{m+e}^{E_{\beta\beta}-m_e}  dE_1 F(Z+2, E_1) F(Z+2, E_2) p_1 E_1 p_2 E_2\ \ \ , 
\ee
 $E_2=E_{\beta\beta} - E_1$, and $F(Z+2, E_{1,2})$ being factors that account for distortion of the electron wave functions in the field of the final state nucleus. 
The NME is given by 
\be
M_0 = \left\langle{\Psi_f} \right\vert \sum_{i,j} \frac{R}{\rho_{ij}} \left[ F_1 {\vec\sigma}_i\cdot {\vec\sigma}_j + F_2 T_{ij}\right]\tau^+_i \tau^+_j\left\vert {\Psi_i}\right\rangle
\ee
where $T_{ij} = 3{\vec\sigma}_i\cdot{\hat\rho}_{ij} {\vec\sigma}_j\cdot{\hat\rho}_{ij} -{\vec\sigma}_i\cdot {\vec\sigma}_j$, $R=r_0 A^{1/3}$,  ${\vec\rho}_{ij}$ is the separation between nucleons $i$ and $j$, and the functions $F_{1,2}(|\vec\rho_{ij} |)$ are given in Ref.~\cite{Prezeau:2003xn}. Note that we have normalized the rate to the conventionally-used factor $G_{0\nu}$ that contains quantities associated with the SM weak interaction, even though the LNV mechanism here involves no SM gauge bosons. The rate (\ref{eq:rate}) is similarly insensitive to the nucleon axial vector coupling $g_A$ and the debate over its \lq\lq quenching" in nuclei\cite{Brown:1988vm,MartinezPinedo:1996vz,Menendez:2011qq,Barea:2013bz,Klos:2013rwa,Engel:2014pha,Ekstrom:2014iya}.

Values for $M_0$ have been computed using the quasiparticle random phase approximation (QRPA) in Ref.~\cite{Faessler:1998qv} for a variety of isotopes. For illustrative purposes, we consider the $0\nu\beta\beta$-decay of $^{76}$Ge, for which the authors of Ref.~\cite{Faessler:1998qv} give $M_0=-1.99$. We emphasize, however, that both the hadronic matching scale $\Lambda_H$ and the NME $M_0$ are presently subject to considerable theoretical uncertainties. In the case of $0\nu\beta\beta$-decay mediated by the exchange of light Majorana neutrinos, for example, NME computations obtained using the nuclear shell model are typically a factor of two smaller than those obtained using QRPA. In order to illustrate the impact of both sources of uncertainty, we show results for two different values the product $M_0\Lambda_H^2$ that differ by a factor of two. 

\begin{figure}
\begin{center}
\hspace*{-0.25cm}
\includegraphics[width=0.475\textwidth]{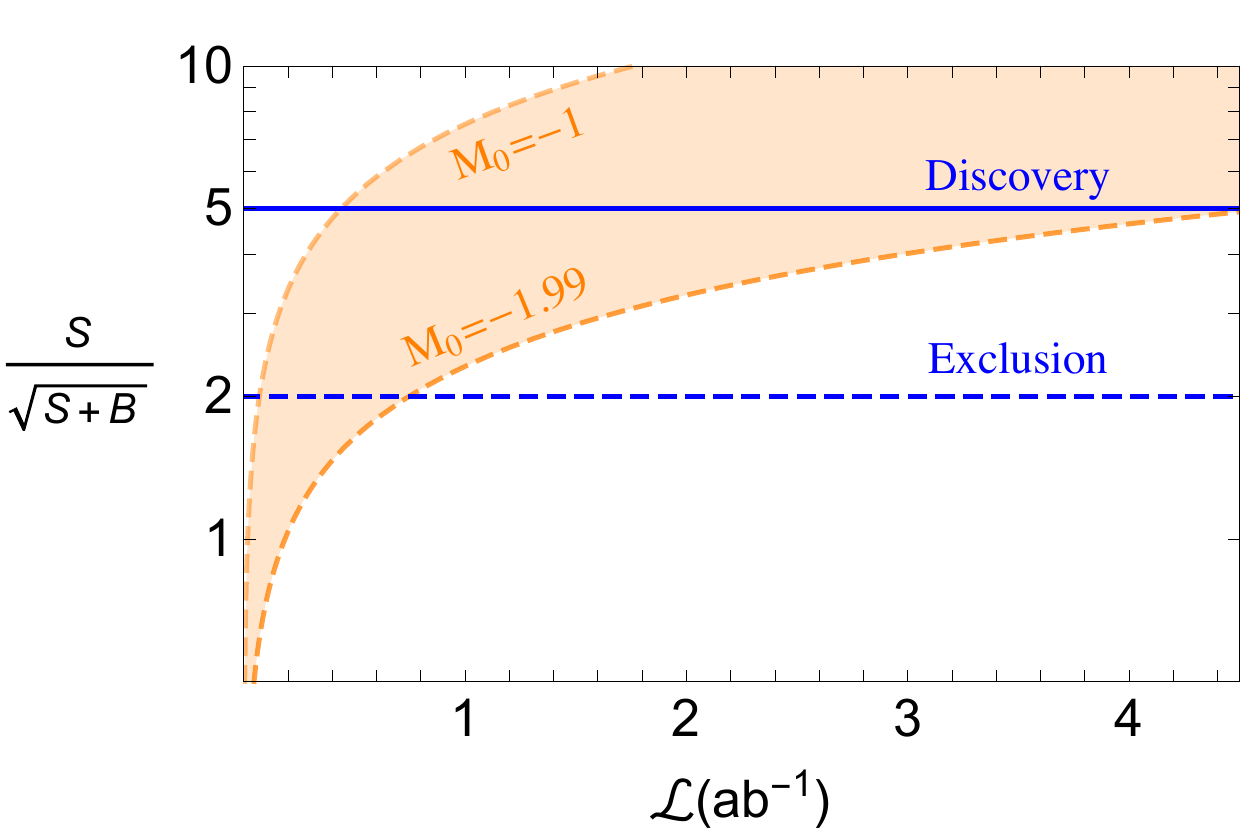} 
\caption{Significance of a LHC e$^-$e$^-$ $+$ di-jet signal as a function of integrated luminosity assuming the maximum 
$C_1/\Lambda^5$ consistent with the GERDA $0\nu\beta\beta$ half-life limit. Upper and lower curves correspond to values of the NME $M_0=-1.0$ and $-1.99$, respectively.
}
\label{fig:lumin}
\end{center}
\end{figure}

To illustrate the present and prospective reach of $0\nu\beta\beta$-decay and LHC searches, we first show in Fig. \ref{fig:lumin} the significance of a possible LHC observation, assuming $C_1/\Lambda^5$ has the maximum value consistent with the present GERDA limit  for $^{76}$Ge ($T_{1/2}<$3$\times$10$^{25}$ yr) as implied by Eq.~(\ref{eq:rate}). We see that non-observation with $\sim 735$ fb$^{-1}$ ($\sim 70$ fb$^{-1}$) would imply exclusion at a level consistent with the present GERDA limit assuming the larger (smaller) value of $M_0\Lambda_H^2$. The corresponding requirement for discovery $S/\sqrt{S+B}\geq 5$ is $\gsim$4.6 ab$^{-1}$ ($\gsim$ 435 fb$^{-1}$). It is striking that a factor of two difference in $M_0\Lambda_H^2$, when translated into an upper bound on $C_1/\Lambda^5$, implies an order of magnitude difference in the luminosity needed for LHC exclusion or discovery.
\begin{figure}
\begin{center}
\hspace*{-0.25cm}
\includegraphics[width=0.475\textwidth]{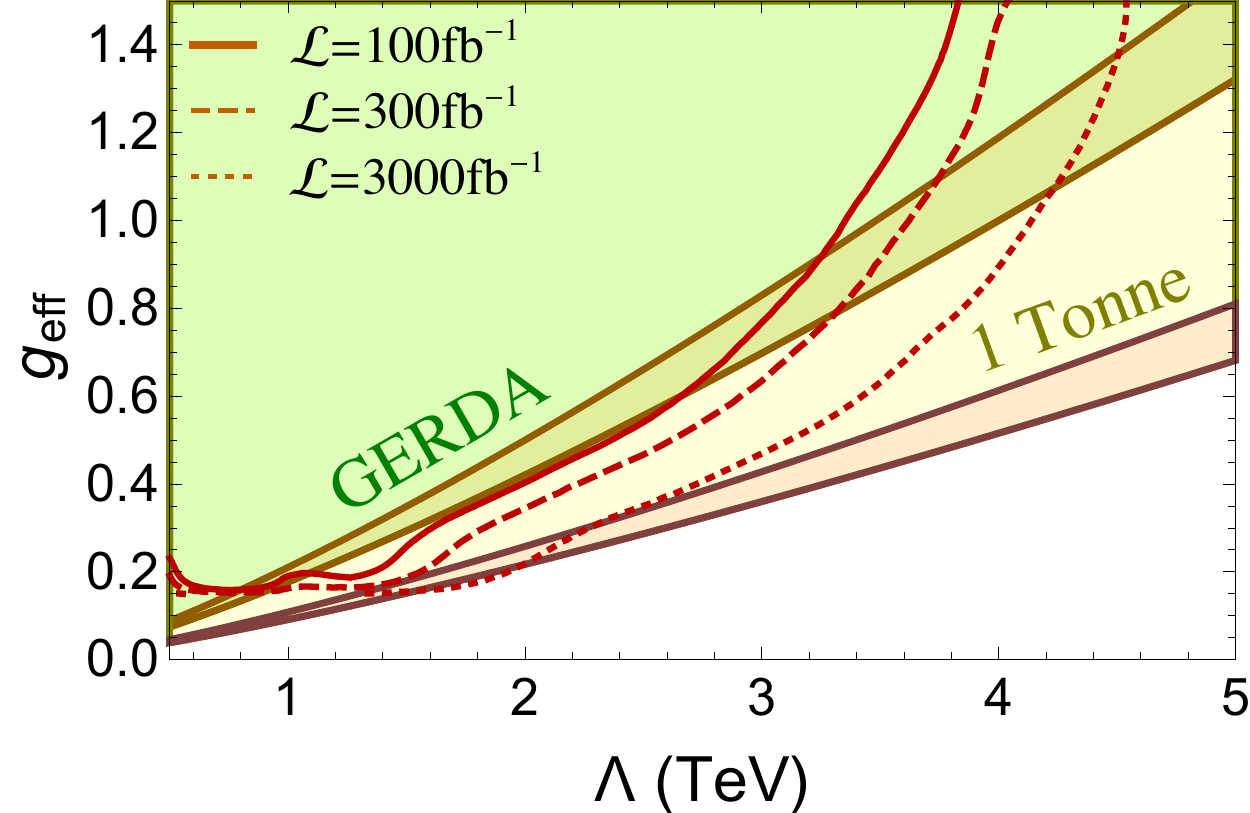} 
\caption{Present and future reach of $0\nu\beta\beta$ and LHC searches for the TeV LNV interaction (\ref{eq:LNV}) as functions of the effective coupling $g_\mathrm{eff}$ and mass scale $\Lambda$ (see text). Present GERDA exclusion and future tonne-scale $0\nu\beta\beta$ sensitivity are indicated by upper and lower shaded regions, respectively. Darker shaded bands indicate impact of varying $M_0\Lambda_H^2$ by a factor of two. LHC exclusion reach for representative integrated luminosities  are indicated by the solid, dashed, and dotted lines.}
\label{fig:exclude}
\end{center}
\end{figure}
\begin{figure}
\begin{center}
\hspace*{-0.25cm}
\includegraphics[width=0.475\textwidth]{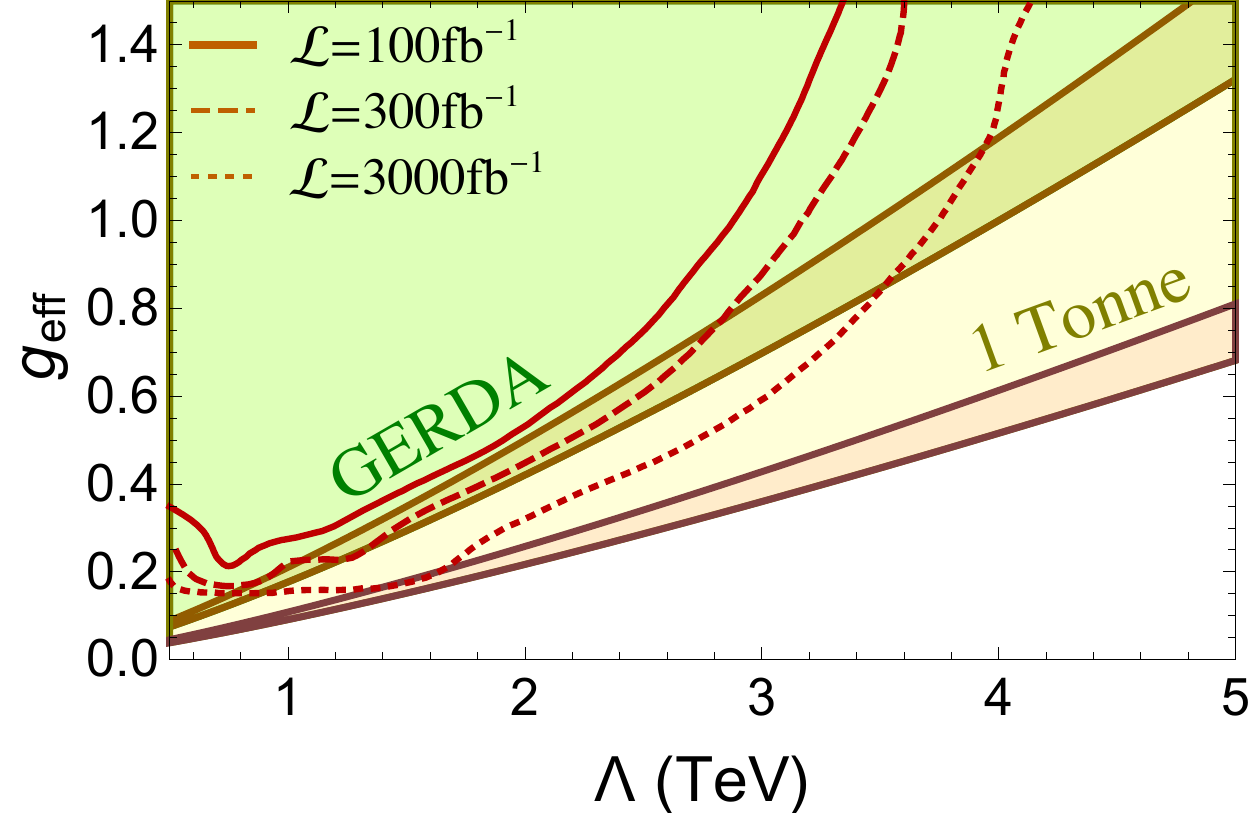} 
\caption{Same as Fig.~\ref{fig:exclude} but giving LHC discovery reach.}
\label{fig:discover}
\end{center}
\end{figure}
The exclusion and discovery reaches for both the LHC and a future, one-ton  $0\nu\beta\beta$-decay as functions of $\Lambda$ and an effective coupling $g_\mathrm{eff} = C_1(\Lambda)^{1/4}$ are shown Figs. \ref{fig:exclude} and \ref{fig:discover}, respectively.  We use a prospective $^{76}$Ge sensitivity of $T_{1/2} = 6\times 10^{27}$ yr\cite{Detwiler:2015}. We also show the present GERDA exclusion for reference. The darker shaded bands at the lower edges of each $0\nu\beta\beta$-decay exclusion and future sensitivity regions indicate the impact of varying $M_0\Lambda_H^2$ by a factor of two. From Fig.~\ref{fig:exclude} we observe that with $\gsim$ 100 fb$^{-1}$ the LHC would begin to extend the present GERDA exclusion for $\Lambda$ in the vicinity of 1.4 TeV for the larger value of $|M_0|\Lambda_H^2$ and for a broader range of masses assuming the smaller value. As indicated by Fig.~ \ref{fig:discover}, the opportunities for discovery with 300 fb$^{-1}$ appear more limited, even under the assumption of the smaller nuclear and hadronic matrix elements. However, the high luminosity phase of the LHC with 3 ab$^{-1}$ could open the possibility for discovery over a range of masses that depends on the value of $M_0\Lambda_H^2$.

From the standpoint of the LHC, this conclusion is not as optimistic as obtained in Refs.~\cite{Helo:2013dla,Helo:2013ika}, as the reach of the tonne-scale  $0\nu\beta\beta$-decay experiments appears to exceed that of the high-luminosity LHC over nearly the entire range of parameter space considered. It is, nevertheless, interesting to compare the prospects for both $0\nu\beta\beta$-decay and the LHC, as observation of a signal in both experiments is possible and would point to the existence of TeV scale LNV interactions. Reducing the $0\nu\beta\beta$-decay nuclear and hadronic matrix element uncertainties, as well as refining the estimates of  jet-fake and charge flip backgrounds at the LHC, would clearly clearly sharpen the implications of this comparison.

\noindent{\it  Acknowledgements} 
We thank B. Balantekin, C. Dallapiccola, J. Detwiler, J.C. Helo, P. Vogel, and S.-L. Wu for helpful discussions. This work was supported in part by U.S. Department of Energy contract  DE-SC0011095.

\bibliographystyle{h-physrev3.bst}
\bibliography{EDMGlobalRefs}

\end{document}